\documentclass[epj]{svjour}
\usepackage{graphics}
\usepackage{amssymb, amsfonts,amsmath}

\def \beq {\begin{equation}}
\def \edq {\end{equation}}

\def \dag {\dagger}
\def \up {\uparrow}
\def \down {\downarrow}

\def \calh {{\cal{H}}}

\def \bef {\begin{framed}}
\def \edf {\end{framed}}
\def \ket {\rangle}
\def \bra {\langle}

\begin{document}
\title{Nonlinear electric and thermoelectric Andreev transport through a hybrid quantum dot coupled to ferromagnetic and superconducting leads}
\titlerunning{Nonlinear Andreev transport through a hybrid quantum dot coupled to F and S leads}
\author{Sun-Yong Hwang\inst{1,2} \and David S\'anchez\inst{1} \and Rosa L\'opez\inst{1}
}                     
%
%
\institute{Institut de F\'{\i}sica Interdisciplin\`aria i Sistemes Complexos IFISC (CSIC-UIB), E-07122 Palma de Mallorca, Spain \and Theoretische Physik, Universit\"at Duisburg-Essen and CENIDE, D-47048 Duisburg, Germany}
\date{Received: date / Revised version: date}
%
\abstract{
We discuss the nonlinear  Andreev current of an interacting quantum dot coupled to spin-polarized and superconducting reservoirs when voltage and temperature biases are applied across the nanostructure. 
Due to the particle-hole symmetry introduced by the superconducting (S) lead, the subgap spin current vanishes identically. Nevertheless, the Andreev charge current depends on the degree of polarization in the ferromagnetic (F) contact since the shift of electrostatic internal potential of the conductor depends on spin orientation of the charge carrier. This spin-dependent potential shift characterizes nonlinear responses in our device. We show how the subgap current versus the bias voltage or temperature difference depends on the lead polarization in two cases, namely (i) S-dominant case, when the dot-superconductor tunneling rate ($\Gamma_R$) is much higher than the ferromagnet-dot tunnel coupling ($\Gamma_L$), and (ii) F-dominant case, when $\Gamma_L\gg \Gamma_R$. For the ferromagnetic dominant case the spin-dependent potential shows a nonmonotonic behavior as the dot level is detuned. Thus the subgap current can also exhibit interesting behaviors such as current rectification and the maximization of thermocurrents with smaller thermal biases when the lead polarization and the quantum dot level are adjusted.
\PACS{
      {74.45.+c}{Proximity effects; Andreev reflection; SN and SNS junctions}   \and
      {74.25.fg}{Thermoelectric effects}	\and
	  {73.23.-b}{Electronic transport in mesoscopic systems}
     } 
\keywords {Andreev reflection, hybrid junctions, thermoelectric effects, nonlinear quantum transport}
} 
\maketitle
\section{Introduction}
Electron-hole conversion processes are dominant at the interface of a superconductor (S) and metallic (N) system~\cite{tin96}. In two consecutive NS interfaces, Andreev bound states are formed by multiple Andreev reflections and a particle current is established for  bias voltages within the subgap energy range. Such subgap states have been recently observed in normal-quantum dot-superconductor (N-QD-S) tunnel experiments~\cite{gra04,Dea10,Pillet10,Schi14,lee14}. These hybrid setups serve as perfect platforms to investigate the interplay between Coulomb interaction and Cooper-pair transport. Whereas Coulomb interaction favors transport of single carriers, superconducting proximity effect leads to transport of electrons as Cooper pairs. The competition between these two interactions yields a variety of phenomena such as the occurrence of Kondo zero-bias anomalies~\cite{Kondo}
or the formation of Yu-Shiba-Rusinov states~\cite{Shiba} and their connection to Majorana quasiparticles~\cite{Majorana,Mourik12}. Furthermore, N-QD-S systems might be relevant in certain quantum information processing setups~\cite{Hof09}. 

Usually, subgap transport through hybrid systems has been investigated by applying solely electric biases. However, quite recently, it has been pointed out that Andreev currents can also be controlled by thermal biases~\cite{andreev}.
It is known that QDs attached to normal contacts generate a thermoelectric voltage $V_\textrm{th}$ in response to a temperature difference $\theta$. Such voltage becomes quite large around a narrow resonance~\cite{mah96} producing high values of the Seebeck coefficient or thermopower $S=-V_\textrm{th}/\theta$~\cite{sta93,dzu93,fah13}. However, the thermoelectric response has been less investigated for interacting hybrid QDs. Since a superconductor is a perfect electric conductor but a poor heat conductor, the thermoelectric conversion should be highly efficient. In fact, when one of the normal contact is replaced by a superconductor the hybrid conductor modifies greatly its thermoelectric properties~\cite{cla95} leading to much higher thermoelectric performances.
Unfortunately, particle-hole symmetry is generally preserved in a superconductor and this leads to a perfect cancellation
of the thermoelectric current due to counterpropagating electron and hole flows. The symmetry can be broken with external magnetic fields or spin-polarized bands~\cite{review1}, leading to sizable values of $S$ as demonstrated
both theoretically~\cite{mac13,oza14,kal14,gia14,mac14,kal15,gia15,hwa16a,bat17,tro17} and experimentally~\cite{kol16,kol16b}.
Another possibility is to drive the junction out of equilibrium beyond the linear response regime. Then,
a cross thermoelectric effect appears in the presence of both electric field and thermal bias~\cite{hwa15}.
In reference~\cite{hwa15}, it was shown that applied thermal bias can increase or decrease the subgap thermocurrent depending on the gate voltages due to this nonlinear cross coupling effect but the discussion was solely for spin symmetric cases.
Here, we combine both effects (spin-polarized injection and nonlinear drivings) to examine how a pure Andreev current
evolves as a function of voltage and temperature biases in a given polarization. Note that reference~\cite{hwa16b} predicts a strong diode behavior that works only
for quasiparticle tunneling. In contrast, we are here interested in the regime where a normal current is transformed into a supercurrent as in the Andreev reflection.
The issue is interesting in view of recent developments in the field of nonlinear quantum thermoelectrics~\cite{review}
and its connection to spin caloritronics (i.e., the production of spin currents with thermal means) in superconductors~\cite{lin16}.

\begin{figure}[t]
\centering
\resizebox{0.4\textwidth}{!}{%
  \includegraphics{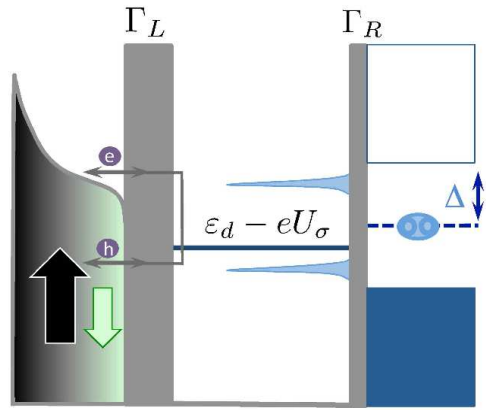}
}
\caption{Energy diagram of a single-level quantum dot coupled to ferromagnetic (left) and superconducting (right) reservoirs. Electron ($e$) and hole ($h$) retroreflections build up Andreev tunneling processes (indicated with horizontal arrows), leading to a finite subgap electric current. The F contact contains different amounts of up ($\uparrow$) and down ($\downarrow$) spins (vertical arrows), giving rise to spin-dependent tunneling rates. Thermal and electrical biases are applied to the left ferromagnetic lead, where a thermally smeared Fermi function is depicted. The superconducting side has a finite gap denoted with $\Delta$ and a Fermi level (dashed energy) common to the F lead. States are filled below the gap. Since we discuss asymmetric couplings ($\Gamma_L\neq \Gamma_R$) we plot the energy barriers (gray areas) with differing widths. The energy level in the quantum dot, $\varepsilon_d$, is renormalized due to the Coulomb potential $U_\sigma$.}
\label{fig:easythrone}       
\end{figure}

Our system comprises an interacting QD attached to a spin-polarized electrode (a ferromagnetic contact, denoted by F) 
and a superconducting reservoir (S), see Figure~\ref{fig:easythrone}.
The F contact is modeled by considering spin-dependent bands with equal chemical potential for both spin orientations
but different densities of states. This leads to spin-dependent tunneling rates for the F-QD coupling around the Fermi level~\cite{ferro}, i.e., $\Gamma_{L\sigma}=\Gamma_L (1+\sigma p)$ with $\sigma=\pm$ being the spin index, and $p$ the polarization degree ($|p|<1$, and $p=0$ for normal contacts).
Since the retroreflected hole after an Andreev conversion process has a spin direction lying opposite to the electron impinging
into the interface, we naturally expect a strong dependence of the Andreev current upon
the polarization of the ferromagnetic contact. However,
remarkably enough, we also find a nontrivial dependence that arises only in the nonlinear regime of transport
due to the spin dependence of the electrostatic internal potential of the quantum dot, which determines the system's screening properties out of equilibrium.  Below, we discuss the details of such dependence in two cases, namely, (i) S-dominant case, when the dot-superconductor tunneling rate ($\Gamma_R$) is much higher than the ferromagnet-dot tunnel coupling ($\Gamma_L$), and (ii) F-dominant case, when $\Gamma_L\gg \Gamma_R$. We anticipate that for the F-dominant case the subgap charge current presents a more dramatic dependence on the lead polarization than that for the S-dominant case.

%
%
%
%
%

\section{Theoretical model}
The Hamiltonian of the system reads
\begin{equation}
\calh=\calh_{L}+\calh_{R}+\calh_{D}+\calh_{T},
\end{equation}
where
\beq
\calh_{L}=\sum_{k\sigma}\varepsilon_{Lk\sigma}c_{Lk\sigma}^{\dag}c_{Lk\sigma}
\edq
describes electrons with momentum $k$, spin $\sigma$ with an internal magnetization $M\sigma$ along a given spin-quantization axis in the left ferromagnetic lead, while
\beq
\calh_{R}=\sum_{p\sigma}\varepsilon_{Rp\sigma}c_{Rp\sigma}^{\dag}c_{Rp\sigma}
	+\sum_{p}\Big[\Delta c_{R,-p\up}^{\dag}c_{Rp\down}^{\dag}+\textrm{H.c.}\Big]
\edq
accounts for the right superconductor. The second term of $\calh_R$ depicts the Cooper pairing inside the energy gap $\Delta$. 
In the QD Hamiltonian
\beq
\calh_{D}=\sum_{\sigma}(\varepsilon_{d}-eU_\sigma)d_{\sigma}^{\dag}d_{\sigma},
\edq
the dot energy level $\varepsilon_{d}$ is renormalized by the spin-de\-pend\-ent internal potential $U_\sigma$ that describes the effect of electron-electron repulsion in the QD within a mean-field approximation. This approach is valid for metallic dots with good screening properties~\cite{bro05}.
We stress that $U_\sigma$ is a nonequilibrium screening potential with its dependence on voltage and temperature biases being determined
by spin-dependent characteristic potentials $u_{L\sigma}$ and $z_{L\sigma}$ as discussed below. These quantities can be found by solving
a simple electrostatic model [the capacitance equation given by equation~\eqref{delta_rho}], which allows us to work with fully analytic expressions
while at the same time capturing the essential physics (when strong correlations can be neglected). Another simplification is to assume a spatially homogeneous potential inside the dot, which qualitatively yields reliable results~\cite{pil02}.
Finally,
\beq
\calh_{T}=\sum_{k\sigma}t_{L\sigma}c_{Lk\sigma}^{\dag}d_{\sigma}
	+\sum_{p\sigma}t_{R\sigma}c_{Rp\sigma}^{\dag}d_{\sigma}+\textrm{H.c.}
\edq
characterizes the tunneling processes between the dot and F and S leads.

We neglect spin-flip scattering (see, however, reference~\cite{cao04}) and determine the spin-resolved current via the time evolution of spin-$\sigma$ electron number in the left lead, i.e., $I_\sigma=-e\bra\dot{N}_{L\sigma}(t)\ket=-(ie/\hbar)\bra[\calh,N_{L\sigma}]\ket$ with $N_{L\sigma}=\sum_{k}c_{Lk\sigma}^{\dag}c_{Lk\sigma}$.
We are specifically interested in the low bias subgap transport regime where Andreev processes are dominant. We write the Andreev current $I_{A\sigma}$ for each spin,
\beq\label{I_A}
I_{A\sigma}=\frac{e}{h}\int d\varepsilon~T_{A\sigma}(\varepsilon)\big[f_{L}(\varepsilon-eV)-f_{L}(\varepsilon+eV)\big],
\edq
where the Fermi-Dirac distribution is given by $f_{\alpha=L,R}(\varepsilon\pm eV)=\{1+\exp[(\varepsilon\pm eV-E_F)/k_{B}T_{\alpha}]\}^{-1}$, with the electrode temperature $T_\alpha=T+\theta_\alpha$ ($T$: background temperature, $\theta_\alpha$: thermal shift) and the voltage bias $V=V_L-V_R$. We take the common Fermi level as the reference energy ($E_F=0$). 

The Andreev transmission $T_{A\sigma}$ in equation~\eqref{I_A} is not only a function of energy $\varepsilon$ but also a function of the potential response $U_\sigma$, which in turn depends on the applied voltage and thermal bias. $T_{A\sigma}$ can be expressed in terms of the dot retarded Green's functions $G^r_{ij}(\varepsilon)$ ($i,j=1,2,3,4$) in the spin-generalized Nambu space
\beq
{\bf G}^r_{d}(\varepsilon)
=\left(\begin{array}{cc|cc}
G^{r}_{11}(\varepsilon)& G^{r}_{12}(\varepsilon)& 0 & 0\\
G^{r}_{21}(\varepsilon)& G^{r}_{22}(\varepsilon)& 0& 0\\
\hline
0& 0 & G^{r}_{33}(\varepsilon) &G^{r}_{34}(\varepsilon)\\
0&0 &G^{r}_{43}(\varepsilon) & G^{r}_{44}(\varepsilon)\\
\end{array}\right),
\edq
where the first block ($i,j=1,2$) [the second block ($i,j=3,4$)] corresponds to the particle spin-up (spin-down) space, with subscripts 1, 3 referring to electron sectors and 2, 4 denoting hole parts. Note that the whole matrix is block-diagonal since we have ignored spin-flip processes, hence separating the spin spaces. We also remark that, for each spin-$\sigma$ space for a particle, the corresponding hole has an opposite spin $\bar\sigma$. In order to determine the spin-resolved Andreev current, we explicitly write for the subgap energy region ($|\varepsilon|<\Delta$)
\begin{multline}\label{GreenUp}
G^{r}_{12}(\varepsilon)=\bigg[\varepsilon-\widetilde{\varepsilon}_{d\up}+\frac{i\Gamma_L}{2}(1+p)+\frac{\widetilde{\Gamma}_R}{2}
-\frac{\Gamma_R^{2}\Delta^{2}A_1^r(\varepsilon)}{4(\Delta^{2}-\varepsilon^2)}\bigg]^{-1}\\
\times\frac{\Gamma_R\Delta A_1^r(\varepsilon)}{2\sqrt{\Delta^2-\varepsilon^{2}}}\,,
\end{multline}
\begin{multline}\label{GreenDown}
G^{r}_{34}(\varepsilon)=-\bigg[\varepsilon-\widetilde{\varepsilon}_{d\down}+\frac{i\Gamma_L}{2}(1-p)+\frac{\widetilde{\Gamma}_R}{2}
-\frac{\Gamma_R^{2}\Delta^{2}A_2^r(\varepsilon)}{4(\Delta^{2}-\varepsilon^2)}\bigg]^{-1}\\
\times\frac{\Gamma_R\Delta A_2^r(\varepsilon)}{2\sqrt{\Delta^2-\varepsilon^{2}}}\,,
\end{multline}
with
\begin{eqnarray}
A_1^r(\varepsilon)&=&\bigg[\varepsilon+\widetilde{\varepsilon}_{d\down}+\frac{i\Gamma_L}{2}(1-p)+\frac{\widetilde{\Gamma}_R}{2}\bigg]^{-1},\label{A1}\\
A_2^r(\varepsilon)&=&\bigg[\varepsilon+\widetilde{\varepsilon}_{d\up}+\frac{i\Gamma_L}{2}(1+p)+\frac{\widetilde{\Gamma}_R}{2}\bigg]^{-1},\label{A2}
\end{eqnarray}
where $\Gamma_{L\sigma}=\Gamma_L(1+\sigma p)=2\pi|t_{L\sigma}|^2 \sum_k \delta (\varepsilon-\varepsilon_{Lk\sigma})$, $\Gamma_{R\sigma}=\Gamma_R=2\pi|t_{R\sigma}|^2\sum_{p}\delta(\varepsilon-\varepsilon_{Rp\sigma})$ ($\Gamma_L$, $\Gamma_R$ are hybridization widths), and $\widetilde{\Gamma}_R=\Gamma_R\varepsilon/\sqrt{\Delta^2-\varepsilon^2}$. For energies above the gap ($|\varepsilon|>\Delta$), we replace $\Gamma_R\Delta/\sqrt{\Delta^2-\varepsilon^{2}}$ in equations~\eqref{GreenUp} and~\eqref{GreenDown} with $i\Gamma_R\Delta\,\textrm{sgn}\,(\varepsilon)/\sqrt{\varepsilon^2-\Delta^2}$ and take $\widetilde{\Gamma}_R=i\Gamma_R|\varepsilon|/\sqrt{\varepsilon^2-\Delta^2}$.
In general, one has $\widetilde{\varepsilon}_{d\sigma}=\varepsilon_d-eU_{\sigma}$ as illustrated in Figure~\ref{fig:easythrone}. This implies that the dot level becomes renormalized due to the value of the interaction potential $U_{\sigma}$. This potential adjusts itself depending on the charge flow into the dot, as we explain below. As a consequence, the system's Green function and transmission properties will generally depend on interactions, even at the mean-field level.
We assume that the external magnetic field is zero and therefore the Zeeman effect is absent.
This is different from the discussions in references~\cite{hwa16a,hwa16b} where magnetic fields are necessary. Moreover, reference~\cite{hwa16b} deals only with quasiparticle contributions in order to generate strong diode effects while reference~\cite{hwa16a} completely neglects the nonlinear phenomena. In this paper, we consider the pure subgap transport in nonlinear regime without magnetic fields.
Hence, the spin-dependent properties will arise solely from the coupling to the ferromagnetic lead. 
However, nonlinear electrostatic potential becomes spin dependent thus creating effective Zeeman splittings of the quantum dot level. This separates the particle and hole energy levels, which is crucial to obtain sizable thermoelectric effects in hybrid systems.
As shown in equations~\eqref{GreenUp}, \eqref{GreenDown}, \eqref{A1} and \eqref{A2}, the spin-up and spin-down interactions can be coupled although we separate the spin spaces, making the F-QD-S problem nontrivial and interesting. However, the subgap transport strictly preserves the particle-hole symmetry hence prohibiting the spin-polarized current, as we now discuss, whereas Andreev charge current can have nontrivial polarization dependence.

With the aid of the Green's functions, the spin-resolved Andreev transmission becomes
\begin{eqnarray}
&T_{A\up}(\varepsilon)=(1-p^2)\Gamma_{L}^{2}\big|G_{12}^{r}(\varepsilon)\big|^{2}\,,\\
&T_{A\down}(\varepsilon)=(1-p^2)\Gamma_{L}^{2}\big|G_{34}^{r}(\varepsilon)\big|^{2}\,,
\end{eqnarray}
from which [via equation~\eqref{I_A}] one can also define the charge ($I_A^{C}=I_{A\up}+I_{A\down}$) and spin-polarized ($I_A^{S}=I_{A\up}-I_{A\down}$) Andreev currents
\begin{multline}\label{IAC}
I_{A}^{C}=(1-p^2)\frac{e\Gamma_{L}^{2}}{h}\int d\varepsilon~ \Big(\big|G_{12}^{r}(\varepsilon)\big|^{2}+\big|G_{34}^{r}(\varepsilon)\big|^{2}\Big)\\
\times\big[f_{L}(\varepsilon-eV)-f_{L}(\varepsilon+eV)\big]\,,
\end{multline}
\begin{multline}\label{IAS}
I_{A}^{S}=(1-p^2)\frac{e\Gamma_{L}^{2}}{h}\int d\varepsilon~ \Big(\big|G_{12}^{r}(\varepsilon)\big|^{2}-\big|G_{34}^{r}(\varepsilon)\big|^{2}\Big)\\
\times\big[f_{L}(\varepsilon-eV)-f_{L}(\varepsilon+eV)\big]\,.
\end{multline}
For $p=0$, i.e., normal lead, we have $|G_{12}^{r}(\varepsilon)|^{2}=|G_{34}^{r}(\varepsilon)|^{2}$, and the problem exactly reproduces the results of N-QD-S junctions~\cite{hwa15}. For the ferromagnetic contact considered here with $p\ne0$, we have $|G_{12}^{r}(\varepsilon)|^{2}\ne|G_{34}^{r}(\varepsilon)|^{2}$ in general, thus yielding spin-polarized transmission $T_{A}^{S}(\varepsilon)=T_{A\up}(\varepsilon)-T_{A\down}(\varepsilon)\ne0$. Yet, due to the intrinsic particle-hole symmetry in the subgap, we always have $T_{A\up}(\varepsilon)=T_{A\down}(-\varepsilon)$ and therefore the energy integration in equation~\eqref{IAS} always vanishes correspondingly, $\int d\varepsilon T_{A}^{S}(\varepsilon)[f_{L}(\varepsilon-eV)-f_{L}(\varepsilon+eV)]=0$.

We now explain how to include the spin-dependent interactions in our setup. We assume a weakly nonequilibrium situation, thereby one can write the nonequilibrium potential~\cite{san13,jacquod} $\delta U=U-U_{\textrm{eq}}=\sum_{\alpha,\sigma}[u_{\alpha\sigma}V_{\alpha}+z_{\alpha\sigma}\theta_{\alpha}]=\sum_{\sigma}\delta U_\sigma$, where the characterstic potentials (CPs) $u_{\alpha\sigma}=(\partial U_\sigma/\partial V_\alpha)_\textrm{eq}$ and $z_{\alpha\sigma}=(\partial U_\sigma/\partial\theta_\alpha)_\textrm{eq}$ respectively determine the potential shift in a spin-dependent manner with respect to applied voltage bias and temperature difference.
For definiteness, we assume that the superconductor is at equilibrium and cold, i.e., $V_R=\theta_R=0$ and $\theta_L=\theta$, which
is the experimentally relevant situation~\cite{kol16}. Next,
we consider the dot density distribution out of equilibrium, $\delta\rho=\rho-\rho_{\textrm{eq}}=\rho_{\textrm{inj}}+\rho_{\textrm{scr}}=\sum_{\sigma}\delta\rho_\sigma$, where $\rho_{\textrm{inj}}$ and $\rho_{\textrm{scr}}$ are the injected and screened density contributions respectively. Finally, we solve the spin-generalized capacitance equation~\cite{jongsoo}
\begin{eqnarray}\label{delta_rho}
\delta\rho_\sigma
&=&\sum_{\alpha=L,q}\Big(D_{\alpha\sigma}^{q}V_{\alpha}+\widetilde{D}_{\alpha\sigma}^{q}\theta_{\alpha}\Big)-\Pi_\sigma\delta U_\sigma \nonumber\\
&=&\sum_{\sigma'}C_{\sigma\sigma'}(\delta U_{\sigma}-V_{\alpha=L}^{\sigma'})\,,
\end{eqnarray}
where $D_{\alpha\sigma}^{q}$ and $\widetilde{D}_{\alpha\sigma}^{q}$ are the charge and entropic injectivities for each spin $\sigma$ with $q=p,h$ denoting the particle and hole contributions, and $C_{\sigma\sigma'}=(1+\sigma' p)C_{\sigma}/2$ is the capacitance coupling between spin $\sigma$ at the dot and spin $\sigma'$ at the left ferromagnetic lead. Hereafter, we simply put $C_{\up}=C_{\down}=C$ and $V_L^{\up}=V_L^{\down}=V$ neglecting the spin voltage.
The latter approximation is good for ferromagnets with fast spin relaxation processes.
The spin-resolved density can be written with the dot lesser Green's function in each spin space
\begin{eqnarray}
&\delta\rho_\up=-i\int d\varepsilon\Big[G_{11}^{<}(\varepsilon)-G_{11,\textrm{eq}}^{<}(\varepsilon)\Big]\,,\\
&\delta\rho_\down=-i\int d\varepsilon\Big[G_{33}^{<}(\varepsilon)-G_{33,\textrm{eq}}^{<}(\varepsilon)\Big]\,,
\end{eqnarray}
with $G_{11}^{<}=\frac{i\Gamma_{L}}{2\pi}[(1+p)|G_{11}^{r}|^{2}f_{L}(\varepsilon-eV)
	+(1-p)|G_{12}^{r}|^{2}f_{L}(\varepsilon+eV)]
+\frac{i\widetilde{\Gamma}_{R}}{2\pi}f_{R}(|G_{11}^{r}|^{2}
	+|G_{12}^{r}|^{2}-\frac{2\Delta}{|\varepsilon|}\textrm{Re}[G_{11}^{r}(G_{12}^{r})^*])$ and
$G_{33}^{<}=\frac{i\Gamma_{L}}{2\pi}[(1-p)|G_{33}^{r}|^{2}f_{L}(\varepsilon-eV)
	+(1+p)|G_{34}^{r}|^{2}f_{L}(\varepsilon+eV)]
+\frac{i\widetilde{\Gamma}_{R}}{2\pi}f_{R}(|G_{33}^{r}|^{2}
	+|G_{34}^{r}|^{2}+\frac{2\Delta}{|\varepsilon|}\textrm{Re}[G_{33}^{r}(G_{34}^{r})^*])$. The diagonal Green's functions $G_{11}^r$ and $G_{33}^r$ can be identified by the relations $G_{12}^r=G_{11}^r(\Gamma_R\Delta/2\sqrt{\Delta^2-\varepsilon^2})A_1^r$ and $G_{34}^r=-G_{33}^r(\Gamma_R\Delta/2\sqrt{\Delta^2-\varepsilon^2})A_2^r$ in equations~\eqref{GreenUp} and \eqref{GreenDown}.
After solving the capacitance equation [viz. equation~\eqref{delta_rho}], we find the spin-dependent CPs
\begin{multline}\label{eq:uLup}
u_{L\up}=\frac{-e\Gamma_{L}}{C_{\up}+\Pi_\up}\int\frac{d\varepsilon}{2\pi}\big(-\partial_\varepsilon f\big)\Big[(1+p)\big|G_{11}^{r}(\varepsilon)\big|^{2}\\
-(1-p)\big|G_{12}^{r}(\varepsilon)\big|^{2}\Big]_\textrm{eq}+\frac{C_\up}{C_{\up}+\Pi_\up}\,,
\end{multline}
\begin{multline}\label{eq:uLdn}
u_{L\down}=\frac{-e\Gamma_{L}}{C_{\down}+\Pi_\down}\int\frac{d\varepsilon}{2\pi}\big(-\partial_\varepsilon f\big)\Big[(1-p)\big|G_{33}^{r}(\varepsilon)\big|^{2}\\
-(1+p)\big|G_{34}^{r}(\varepsilon)\big|^{2}\Big]_\textrm{eq}+\frac{C_\down}{C_{\down}+\Pi_\down}\,,
\end{multline}
and
\begin{multline}\label{eq:zLup}
z_{L\up}=\frac{-\Gamma_{L}}{C_{\up}+\Pi_\up}\int\frac{d\varepsilon}{2\pi}\frac{\varepsilon-E_{F}}{T}\big(-\partial_\varepsilon f\big)\\
\times\Big[(1+p)\big|G_{11}^{r}(\varepsilon)\big|^{2}+(1-p)\big|G_{12}^{r}(\varepsilon)\big|^{2}\Big]_\textrm{eq}\,,
\end{multline}
\begin{multline}\label{eq:zLdn}
z_{L\down}=\frac{-\Gamma_{L}}{C_{\down}+\Pi_\down}\int\frac{d\varepsilon}{2\pi}\frac{\varepsilon-E_{F}}{T}\big(-\partial_\varepsilon f\big)\\
\times\Big[(1-p)\big|G_{33}^{r}(\varepsilon)\big|^{2}+(1+p)\big|G_{34}^{r}(\varepsilon)\big|^{2}\Big]_\textrm{eq}\,,
\end{multline}
where the Lindhard functions~\cite{buttiker} for each spin and particle ($p$) or hole ($h$) are given by ($\Pi_{\sigma}=\Pi_{\sigma}^{p}+\Pi_{\sigma}^h$)
\begin{multline}
\Pi^{p}_{\up}=\int\frac{d\varepsilon}{2\pi}f_{\textrm{eq}}\Bigg[\Gamma_{L}(1+p)
\frac{\delta\big|G_{11}^{r}(\varepsilon)\big|^{2}}{\delta U_\up}\\
+\widetilde{\Gamma}_{R}
\bigg(\frac{\delta\big|G_{11}^{r}(\varepsilon)\big|^{2}}{\delta U_\up}
-\frac{\Delta}{|\varepsilon|}\frac{\delta}{\delta U_\up}G_{11}^{r}(\varepsilon)\big[G_{12}^{r}(\varepsilon)\big]^{*}\bigg)\Bigg]_{\textrm{eq}}\,,
\end{multline}
\begin{multline}
\Pi^{h}_{\up}=\int\frac{d\varepsilon}{2\pi}f_{\textrm{eq}}\Bigg[\Gamma_{L}(1-p)
\frac{\delta\big|G_{12}^{r}(\varepsilon)\big|^{2}}{\delta U_\up}\\
+\widetilde{\Gamma}_{R}
\bigg(\frac{\delta\big|G_{12}^{r}(\varepsilon)\big|^{2}}{\delta U_\up}
-\frac{\Delta}{|\varepsilon|}\frac{\delta}{\delta U_\up}G_{12}^{r}(\varepsilon)\big[G_{11}^{r}(\varepsilon)\big]^{*}\bigg)\Bigg]_{\textrm{eq}}\,,
\end{multline}
\begin{multline}
\Pi^{p}_{\down}=\int\frac{d\varepsilon}{2\pi}f_{\textrm{eq}}\Bigg[\Gamma_{L}(1-p)
\frac{\delta\big|G_{33}^{r}(\varepsilon)\big|^{2}}{\delta U_\down}\\
+\widetilde{\Gamma}_{R}
\bigg(\frac{\delta\big|G_{33}^{r}(\varepsilon)\big|^{2}}{\delta U_\down}
+\frac{\Delta}{|\varepsilon|}\frac{\delta}{\delta U_\down}G_{33}^{r}(\varepsilon)\big[G_{34}^{r}(\varepsilon)\big]^{*}\bigg)\Bigg]_{\textrm{eq}}\,,
\end{multline}
\begin{multline}
\Pi^{h}_{\down}=\int\frac{d\varepsilon}{2\pi}f_{\textrm{eq}}\Bigg[\Gamma_{L}(1+p)
\frac{\delta\big|G_{34}^{r}(\varepsilon)\big|^{2}}{\delta U_\down}\\
+\widetilde{\Gamma}_{R}
\bigg(\frac{\delta\big|G_{34}^{r}(\varepsilon)\big|^{2}}{\delta U_\down}
+\frac{\Delta}{|\varepsilon|}\frac{\delta}{\delta U_\down}G_{34}^{r}(\varepsilon)\big[G_{33}^{r}(\varepsilon)\big]^{*}\bigg)\Bigg]_{\textrm{eq}}\,,
\end{multline}
which describe spin-dependent screening effects.
We are now in a position to calculate the Andreev current.

\begin{figure}[t]
\resizebox{0.5\textwidth}{!}{%
  \includegraphics{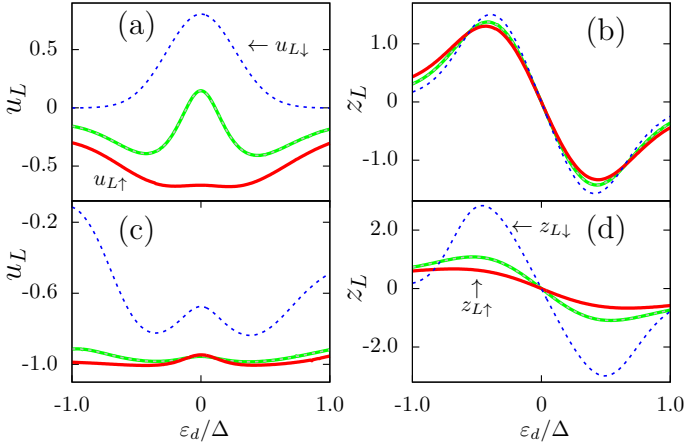}
}
  \caption{Characteristic potentials $u_{L\sigma}$, $z_{L\sigma}$ as a function of the dot level position $\varepsilon_d$ for $\Gamma_L=0.1\Delta$, $\Gamma_R=0.5\Delta$ [(a),(b)] and $\Gamma_L=0.5\Delta$, $\Gamma_R=0.1\Delta$ [(c),(d)]. Green and grey (red and blue) lines indicate $u_{L\up}$ or $z_{L\up}$ and $u_{L\down}$ or $z_{L\down}$ with $p=0$ ($p=0.9$).
Green and gray lines overlap since no spin separation is expected for $p=0$.  
  We use Fermi level $E_F=0$, capacitive couplings $C_\up=C_\down=0$, and background temperature $k_BT=0.1\Delta$.}
  \label{fig:CPs}
\end{figure}

\section{Results and discussions}
With the solutions in the previous section that clearly include spin-generalized interaction effects, we next discuss the polarization dependence of the subgap transport, i.e., $I_A=I_A^C$ in equation~\eqref{IAC}. As aforementioned, the spin current $I_{A}^S$ in equation~\eqref{IAS} is identically zero in our setup due to the inherent particle-hole symmetry. However, one might envisage spin-polarized transport of the crossed Andreev current in multiterminal setups~\cite{tro14,tro15,fut10,wys12,mic15} or the spin-polarized energy flow in the subgap regime.
We here focus only on the spin-unpolarized Andreev current $I_A$ and discuss the nontrivial dependence on the ferromagnetic lead polarization.

Since the interesting effects stem from the competition between F and S leads, it will be meaningful to separately discuss the cases where the coupling strengths are (i) S-dominant ($\Gamma_L\ll\Gamma_R$) and (ii) F-dominant ($\Gamma_L\gg\Gamma_R$) with several cases of polarization described by the parameter $p$. In addition, we explore the gate-dependence of the Andreev current for a given polarization. Below, we use $\Gamma_L=0.1\Delta$, $\Gamma_R=0.5\Delta$ for S-dominant case, and $\Gamma_L=0.5\Delta$, $\Gamma_R=0.1\Delta$ for F-dominant case.

Firstly, we discuss the spin-dependent CPs with a non-zero polarization $p$. Figure~\ref{fig:CPs} displays the solutions of equations~\eqref{eq:uLup}, \eqref{eq:uLdn}, \eqref{eq:zLup} and \eqref{eq:zLdn} as a function of $\varepsilon_d$ with $p=0$ and $p=0.9$, for S-dominant [(a),(b)] and F-dominant [(c),(d)] cases.
Nonzero polarization generally separates CPs, i.e., $u_{L\up}\ne u_{L\down}$ and $z_{L\up}\ne z_{L\down}$. Moreover, this spin separation is highly nonmonotonic depending on the gate potential. In Figure~\ref{fig:CPs}(a), for S-dominant case, $u_{L\up}$ and $u_{L\down}$ become separated as $p$ is increased away from the normal (unpolarized) case where $u_{L\up}=u_{L\down}$ and finally $u_{L\down}>0$ and $u_{L\up}<0$ for a high enough polarization here shown with $p=0.9$.
Interestingly, for F-dominant case as shown in Figure~\ref{fig:CPs}(c), only the minority spin CP $u_{L\down}$ drastically changes departing from the unpolarized curve while the majority one $u_{L\up}$ is rather robust with an increasing spin-up population in the F-lead. Hence, for a high $p$ with $\Gamma_{L\up}\gg\Gamma_{L\down}$, the applied voltage shifts the spin-up dot level as $\varepsilon_{d\up}=\varepsilon_d-u_{L\up}eV\approx\varepsilon_d+eV$, strengthening the charge neutrality condition observed in N-QD-S case~\cite{hwa15}. Importantly, this insensitivity of the majority spin CP $u_{L\up}$ to $p$ is related to the huge rectification of the Andreev current displayed in Figure~\ref{fig:Fdom}(c) which is also insensitive to $p$ for the positively applied voltage bias.

The effect of $z_L$ polarization for S-dominant case is very small, i.e., $z_{L\up}\approx z_{L\down}$, as shown in Figure~\ref{fig:CPs}(b). However, when the coupling to ferromagnetic lead is much stronger, the polarization effect can become pronounced as displayed in Figure~\ref{fig:CPs}(d). Also, as discussed in N-QD-S case~\cite{hwa15}, $z_{L\up}=z_{L\down}=0$ is always satisfied exactly at the symmetric point $\varepsilon_d=E_F$. Intriguingly, by applying the gate potential, one can find the condition $z_{L\up}=z_{L\down}$ at some $\varepsilon_d\ne0$, after which $z_{L\up}>z_{L\down}$. This is indeed distinct from the voltage driven CP, where we always have $u_{L\up}<u_{L\down}$ with $p$, e.g., $-eu_{L\up}V>-eu_{L\down}V$ for positive voltage. Hence, renormalized dot level $\varepsilon_{d\sigma}=\varepsilon_d-eu_{L\sigma}V$ in nonequilibrium potential always gives $\varepsilon_{d\up}>\varepsilon_{d\down}$ as $p$ is increased. As a consequence, the results shown here for $z_{L}$ suggest that the level may shift to a direction opposite to the voltage driving case.

We observe that a left-right asymmetry with respect to $\varepsilon_d$ appears only with the nonzero polarization $p$ and is more visible in the F-dominant case as shown in Figures~\ref{fig:CPs}(c) and \ref{fig:CPs}(d). We remark, however, that even in the S-dominant case [Figures~\ref{fig:CPs}(a) and ~\ref{fig:CPs}(b)] this asymmetry still exists although it is small. When the quantum dot is attached to a normal lead, i.e., $p=0$, the asymmetry disappears as shown with green and gray lines. It should be mentioned that this is a purely nonlinear effect since the characteristic potentials $u_L$ and $z_L$ describing the nonlinear response are zero in linear response. Hence, these nonlinear potentials act as effective Zeeman splittings as they shift the bare dot level, e.g., $\varepsilon_{d\uparrow}\to \varepsilon_d -u_{L\uparrow}$, etc., when we apply voltage or temperature. The general symmetry is preserved with a reversal of both polarization and the spin directions:  $u_{L\uparrow} (p)= u_{L\downarrow} (-p)$ and $z_{L\uparrow} (p)=z_{L\downarrow} (-p)$. 

\begin{figure}[htbp]
\resizebox{0.5\textwidth}{!}{%
  \includegraphics{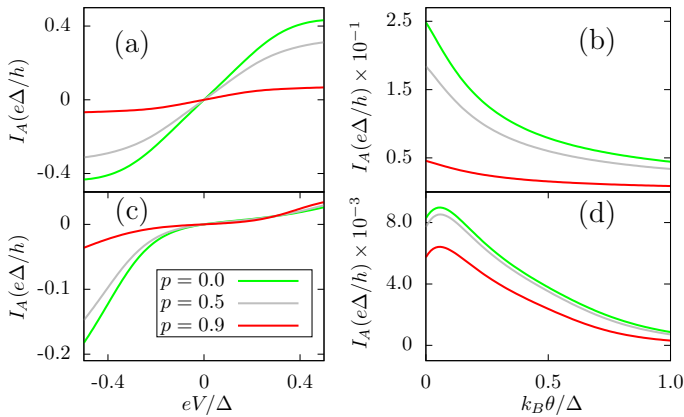}
}
  \caption{Andreev current $I_A$ versus (a), (c) voltage $eV$ with temperature bias $k_B\theta=0$ (b), (d) $k_B\theta$ with $eV=0.2\Delta$, for S-dominant case where $\Gamma_L=0.1\Delta$, $\Gamma_R=0.5\Delta$ with several spin polarization values $p=0, 0.5, 0.9$. In (a),(b) [(c),(d)] we use $\varepsilon_d=0$ [$\varepsilon_d=0.5\Delta$]. Green, gray, and red lines indicate the cases of $p=0$, 0.5, 0.9, respectively. Parameters: $E_F=0$, $C_\up=C_\down=0$, and $k_BT=0.1\Delta$.}
  \label{fig:Sdom}
\end{figure}

In Figure~\ref{fig:Sdom}, we plot the Andreev current when the coupling to the S-lead is dominant as a function of bias voltage in (a) and (c), and as that of temperature difference in (b) and (d). For the latter case we need to apply a nonzero voltage (e.g., $eV=0.2\Delta$); otherwise Andreev thermocurrent is always zero even if the gate potential is applied due to the subgap particle-hole symmetry~\cite{hwa16a}. In Figures~\ref{fig:Sdom}(a) and \ref{fig:Sdom}(b) where $\varepsilon_d=0$ is at the symmetric point, the $p$-dependence is trivial mainly due to the prefactor $1-p^2$ in equation~\eqref{IAC}. However, this becomes very different when we apply a gate potential to shift the dot level up to $\varepsilon_d=0.5\Delta$, as shown in Figures~\ref{fig:Sdom}(c) and \ref{fig:Sdom}(d). First, the $I_A-V$ curve shows a strong rectification for $\varepsilon_d=0.5\Delta$ [Figure~\ref{fig:Sdom}(c)]. As temperature difference increases, $I_A$ generally tends to zero monotonically for $\varepsilon_d=0$, which, however, can be maximized with respect to bias configurations (here the maximum point is shown at $eV=0.2\Delta$ and $k_B\theta=0.06\Delta$) with a detuned dot level $\varepsilon_d=0.5\Delta$ [Figure~\ref{fig:Sdom}(d)]. This is the case for any given $p$. Thus, the current $I_A$ can be tuned with biases and gate potential in a given polarization.

When the coupling to the F-lead is dominant as shown in Figure~\ref{fig:Fdom}, the $p$-dependence becomes highly nonmonotonic. In Figure~\ref{fig:Fdom}(a), $I_A-V$ curve is rather robust to an increasing $p$ up to $p=0.5$. But as the F-lead is more polarized, $I_A$ suddenly increases as voltage is applied instead of exhibiting $1-p^2$ decrease as expected for S-dominant case [compare with Figure~\ref{fig:Sdom}(a)]. This happens in the nonlinear regime of transport (but still within the superconducting gap) with an interesting interplay between ferromagnetism and interaction effect. This is the result for $\varepsilon_d=0$. Figure~\ref{fig:Fdom}(c) shows the results with applied gate potential $\varepsilon_d=0.5\Delta$, in which the Andreev current is highly rectified in any polarization $p$. The thermal driving case at the symmetric point $\varepsilon_d=0$ is shown in Figure~\ref{fig:Fdom}(b). Here, $I_A-\theta$ curve shows an insensitivity above a certain temperature difference with varying $p$. This can be understood since CPs $z_{L\up}=z_{L\down}=0$ for $\varepsilon_d=0$ as shown in Figure~\ref{fig:CPs}(d). However, when we detune the dot level to $\varepsilon_d=0.5\Delta$ [Figure~\ref{fig:Fdom}(d)], it becomes highly sensitive to the lead polarization because the potential shift is strongly spin dependent, see Figure~\ref{fig:CPs}(d). It should be emphasized that the F-dominant case shows nonmonotonic $I_A-\theta$ curve as a function of $p$, compared to simple $1-p^2$ dependence in Figures~\ref{fig:Sdom}(b) and \ref{fig:Sdom}(d).
For S-dominant case, the polarization dependence is rather monotonic since the difference between $z_{L\up}$ and $z_{L\down}$ is not so pronounced even with a high polarization $p=0.9$ as displayed in Figure~\ref{fig:CPs}(b). This holds for a broad range of gate potentials.

\begin{figure}[htbp]
\resizebox{0.5\textwidth}{!}{%
  \includegraphics{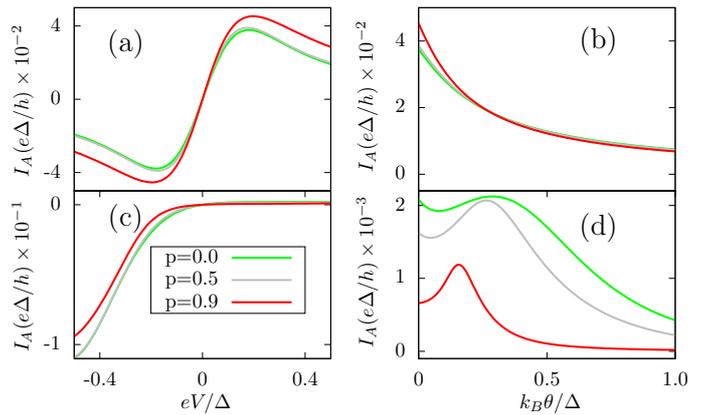}
}
  \caption{Andreev current $I_A$ versus (a), (c) voltage $eV$ with temperature bias $k_B\theta=0$ (b), (d) $k_B\theta$ with $eV=0.2\Delta$, for F-dominant case where $\Gamma_L=0.5\Delta$, $\Gamma_R=0.1\Delta$ with several spin polarization values $p=0, 0.5, 0.9$. In (a),(b) [(c),(d)] we use $\varepsilon_d=0$ [$\varepsilon_d=0.5\Delta$]. Green, gray, and red lines indicate the cases of $p=0$, 0.5, 0.9, respectively. Parameters: $E_F=0$, $C_\up=C_\down=0$, and $k_BT=0.1\Delta$.}
  \label{fig:Fdom}
\end{figure}

\section{Conclusions}
In closing, we have investigated the thermoelectric properties of the Andreev current in a F-QD-S hybrid system. Due to the presence of the ferromagnetic electrode, electrons possess different tunneling rates depending on their spin orientations.
Therefore, the Andreev current becomes a function of the ferromagnetic degree of polarization when the driving force is either electrical or thermal. 

Our main findings can be understood from the spin-dependent characteristic potentials shown in Figure 2 with specific examples for the Andreev current displayed in Figures 3 and 4. We have derived analytic expressions for these characteristic potentials in a spin-generalized manner assuming the quantum dot as a single capacitor. We have emphasized that the distinction between Figures 2(b) and 2(d) is important as the lowest nonvanishing Andreev thermocurrent can only be described by these characteristic potentials. Had it not been for the mixed thermoelectric response quantified by $z_{L\up}$ and $z_{L\down}$ with nonzero voltage bias, Andreev thermocurrents in Figures 3 and 4 would have always been zero irrespective of ferromagnetic lead polarization and also of applied gate potential. Note that a minute polarization dependence in S-dominant case [Figure 2(b)] in stark contrast to the oppsite case [Figure 2(d)]. We have also put an equal emphasis on the difference between Figures 2(a) and 2(c). In F-dominant case [Figure 2(c)], potential shift of the minority spin component is drastic as the gate potential is applied while the majority one maintains the charge neutrality. However, effective level splittings due to spin polarization in S-dominant case [Figure 2(a)] are rather monotonic albeit unequal splittings from the unpolarized positions.

As a consequence, the dependence of subgap transport on the lead polarization has a stronger impact in the ferromagnet dominant case with $\Gamma_L\gg\Gamma_R$ than the opposite limit.
This is an interesting finding that is visible only in the nonlinear regime of transport since interactions are crucial beyond linear response.	
Our results thus show how the Andreev current can be modified by thermal and electrical biases and how these dependences can be tuned by varying the ferromagnetic lead polarization, which might have interesting consequences for the design of future spin caloritronic hybrid devices.

%
\begin{acknowledgement}
This research was supported by MINECO under Grant No.\ FIS2014-52564, the Ministry of Innovation NRW and the Korean NRF under Grant No. 2014R1A6A3A03059105.
\end{acknowledgement}

\section*{Author contribution statement}
All authors analyzed the results and wrote the paper.

%

\end{document}